\documentclass{elsart}

\usepackage{amssymb}
\usepackage{epsfig}
\usepackage{graphics}
\usepackage{longtable}
\begin{document}

\begin{frontmatter}

\title{Event-by-event fluctuations of the mean transverse momentum 
in 40, 80, and 158~A~GeV/$c$ Pb-Au collisions}

\begin{center}
\author{CERES Collaboration}
\end{center}

\author[a]{D.~Adamov\'{a}},
\author[b]{G.~Agakichiev},
\author[c]{H.~Appelsh\"{a}user},
\author[d]{V.~Belaga},
\author[b]{P.~Braun-Munzinger},
\author[c]{R.~Campagnolo},
\author[b]{A.~Castillo},
\author[e]{A.~Cherlin},
\author[c]{S.~Damjanovi\'{c}},
\author[c]{T.~Dietel},
\author[c]{L.~Dietrich},
\author[f]{A.~Drees},
\author[c]{S.~Esumi},
\author[c]{K.~Filimonov},
\author[d]{K.~Fomenko},
\author[e]{Z.~Fraenkel},
\author[b]{C.~Garabatos},
\author[c]{P.~Gl\"{a}ssel},
\author[b]{G.~Hering},
\author[b]{J.~Holeczek},
\author[a]{V.~Kushpil},
\author[g]{B.~Lenkeit},
\author[c]{W.~Ludolphs},
\author[b]{A.~Maas},
\author[b]{A.~Mar\'{\i}n},
\author[c]{J.~Milo\v{s}evi\'{c}},
\author[e]{A.~Milov},
\author[b]{D.~Mi\'{s}kowiec},
\author[g]{L.~Musa},
\author[d]{Yu.~Panebrattsev},
\author[d]{O.~Petchenova},
\author[c]{V.~Petr\'{a}\v{c}ek},
\author[g]{A.~Pfeiffer},
\author[b]{J.~Rak},
\author[e]{I.~Ravinovich},
\author[h]{P.~Rehak},
\author[c]{M.~Richter},
\author[b]{H.~Sako},
\author[c]{W.~Schmitz},
\author[g]{J.~Schukraft},
\author[b]{S.~Sedykh},
\author[c]{W.~Seipp},
\author[b]{A.~Sharma},
\author[d]{S.~Shimansky},
\author[c]{J.~Sl\'{\i}vov\'a},
\author[c]{H.~J.~Specht},
\author[c]{J.~Stachel},
\author[a]{M.~\v{S}umbera},
\author[c]{H.~Tilsner},
\author[e]{I.~Tserruya},
\author[b,i]{J.~P.~Wessels},
\author[c]{T.~Wienold},
\author[c]{B.~Windelband},
\author[j]{J.~P.~Wurm},
\author[e]{W.~Xie},
\author[c]{S.~Yurevich},
\author[d]{V.~Yurevich}
\address[a]{Nuclear Physics Institute ASCR, 25068 \v{R}e\v{z}, Czech Republic}
\address[b]{Gesellschaft f\"ur Schwerionenforschung (GSI), 64291 Darmstadt, Germany}
\address[c]{Physikalisches Institut der Universit\"{a}t Heidelberg, 69120 Heidelberg, Germany}
\address[d]{Joint Institute for Nuclear Research, 141980 Dubna, Russia}
\address[e]{Weizmann Institute, Rehovot 76100, Israel}
\address[f]{Department of Physics and Astronomy, State University of New York-Stony Brook, Stony Brook,
New York 11794-3800, U.S.A.}
\address[g]{CERN, 1211 Geneva 23, Switzerland}
\address[h]{Brookhaven National Laboratory, Upton, New York 11973-5000, U.S.A.}
\address[i]{Institut f\"ur Kernphysik der Universit\"at M\"unster, 48149 M\"unster, Germany}
\address[j]{Max-Planck-Institut f\"{u}r Kernphysik, 69117 Heidelberg, Germany}

\author{}

\address{}

~

\begin{abstract}
Measurements of event-by-event fluctuations of the mean transverse momentum 
in Pb-Au collisions at
40, 80, and 158~A~GeV/$c$ are presented.
A significant excess of mean $p_T$ fluctuations at mid-rapidity is observed over the
expectation from statistically independent particle 
emission.
The results are somewhat smaller than
recent measurements at RHIC. 
A possible non-monotonic 
behaviour of the mean $p_T$ fluctuations as function
of collision energy, which may have indicated
that the system has passed
the critical point of the QCD phase diagram in the range
of $\mu_B$ under investigation, has not been observed.
The centrality dependence of mean $p_T$ fluctuations in Pb-Au
is consistent with an extrapolation from $pp$ collisions 
assuming that the non-statistical fluctuations scale with multiplicity.
The results are compared to calculations by the {\sc rqmd} and
{\sc urqmd} event generators.
\end{abstract}

\end{frontmatter}

\section{Introduction}

The investigation of high energy nucleus-nucleus collisions provides a unique
tool to study the properties of hot and dense nuclear matter. 
The motivation largely comes from QCD calculations, particularly on the
lattice, which predict, at sufficiently high temperatures and densities,
a transition from hadronic matter to a plasma
of deconfined quarks and gluons. 
In this phase chiral symmetry is also restored.
For the case of vanishing net baryon density
(or baryon chemical potential $\mu_B$), state of the
art lattice calculations 
suggest that the QCD phase transition occurs at a critical temperature 
of $T_c$=175$\pm$8~MeV~\cite{Karsch:2001}.
However, a full exploration of the QCD phase diagram in the $T$-$\mu_B$-plane
is desired to obtain insight into the mechanism of deconfinement and 
chiral symmetry restoration, as well as the properties of astrophysical 
objects such as neutron stars.
At finite net baryon chemical potential $\mu_B$, lattice calculations
run into technical limitations, which have been partially
overcome recently~\cite{Fodor:2002,Allton:2002}, showing how $T_c$ drops 
with increasing $\mu_B$. 
Complex structures of the QCD phase diagram have been unveiled,
such as the possible existence
of a critical point of the phase boundary, and a color-superconducting
phase at very high baryon density and low 
temperature~\cite{Fodor:2002,Berges:1998,Halasz:1998,AlfordRapp:1998}.

\newpage

Most theoretical studies of the phase transition imply that it is
second order or a continuous but rapid cross-over, 
at least for $\mu_B$ values less than 
500~MeV~\cite{Karsch:2001,Fodor:2002,Allton:2002}.
Passage of a system through a second order transition or close to a critical
point may lead to critical phenomena, long-range correlations and large
fluctuations. 
The study of event-by-event fluctuations therefore provides a novel probe
to explore the QCD phase diagram, searching for the quark-gluon plasma (QGP) 
and the QCD critical point. Such measurements became possible 
with large acceptance experiments at SPS and RHIC, where the high multiplicity 
of charged particles produced in collisions of lead and gold nuclei allows a 
precise determination
of global observables on an event-by-event basis. Pioneering experimental studies have 
been performed by the NA49 experiment at the SPS, on fluctuations
of the mean transverse momentum $p_{T}$~\cite{NA49:1999} and of the $K/\pi$
ratio~\cite{NA49:2001} in central Pb-Pb collisions at 158~A~GeV/$c$.

Mean $p_{T}$ fluctuations have been investigated in $pp$ collisions at the 
ISR~\cite{braune_pp}.
These data may be used as a reference to study the degree of thermal equilibration
in nucleus-nucleus collisions~\cite{Gazdzicki:1992,Gazdzicki:1999}. 
On the other hand, an enhancement of mean $p_{T}$ fluctuations might be
related to non-trivial effects showing up specifically in nuclear reactions.
It was predicted that mean $p_{T}$ fluctuations can be enhanced
if the system passes through the QCD critical point, 
where long
wave length fluctuations of the sigma field develop, leading to
fluctuations of pions through the strong $\sigma$-$\pi$-$\pi$
coupling~\cite{Stephanov:1998,Stephanov:1999}.
Decay of the Polyakov loop~\cite{Dumitru:2001} and random color
fluctuations~\cite{Mrowczynski:1993} are also proposed as
possible mechanisms to enhance momentum fluctuations 
when the system passes the phase boundary from the QGP to the
hadronic phase.
At high baryon chemical potential ($\mu_B\approx 3T$), 
the possible occurence of large baryon number density fluctuations
has been discussed~\cite{Swansea:2003}. 

The NA49 experiment observed non-statistical 
mean $p_{T}$ fluctuations consistent with zero~\cite{NA49:1999}
at forward rapidity in 158~A~GeV/$c$ Pb-Pb collisions at the SPS.
This result was explained by 
a small positive contribution from Bose-Einstein correlations, 
which is quantitatively compensated by a negative
contribution arising from the finite two-track resolution of the 
detector. No indications for additional non-statistical fluctuations
were observed at forward rapidity. However, the complex structure
of the QCD phase diagram requires a detailed study
of event-by-event fluctuations in the central rapidity region and
at all available beam energies. In particular, the 
data taken at lower SPS energies allow for
an investigation of event-by-event fluctuations at larger $\mu_B$.
The observation of a possible non-monotonic behaviour of the fluctuation strength
as function of $\mu_B$ may indicate the location of the critical point in the
QCD phase diagram.

In this paper, we present measurements of event-by-event fluctuations of the 
mean transverse momentum $p_{T}$ of charged particles
produced near mid-rapidity in Pb-Au collisions at 40, 80, and 158~A~GeV/$c$.

%
%
%
%

\section{Experimental setup}
\label{Sec:ceres-setup}
\begin{figure}[t]
\begin{center}
\mbox{\epsfig{file=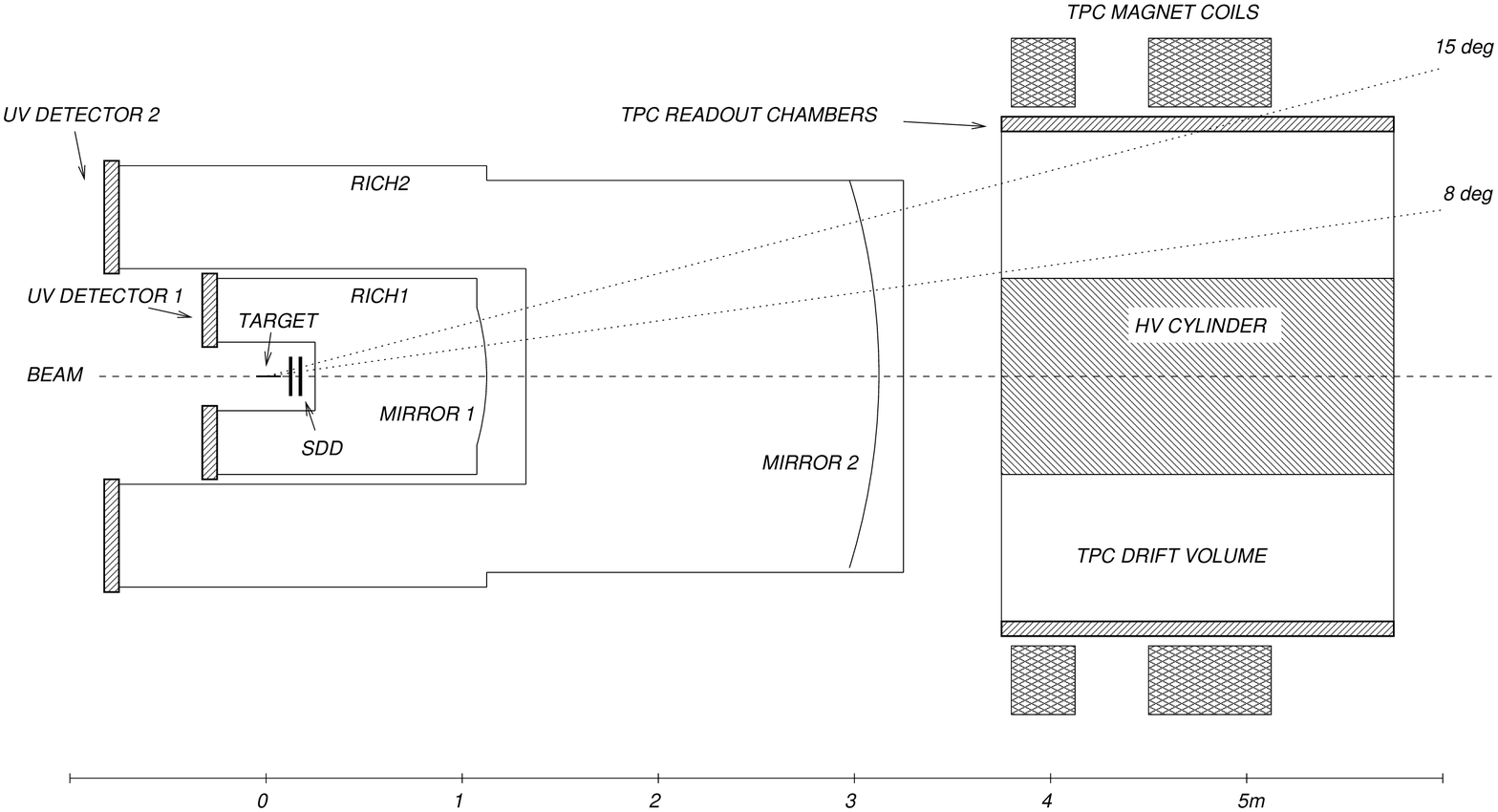,height=7cm}}
\end{center}
\caption{The CERES experiment.}
\label{Fig:setup}
\end{figure}

The CERES spectrometer at the CERN-SPS was originally
conceived for the measurement of low mass $e^+e^-$-pairs produced in proton and
heavy-ion induced collisions with nuclei~\cite{CERES:pa-1998}. 
This version of CERES consisted of two radial silicon drift detectors 
(SDD's) and two ring-imaging Cherenkov
counters (RICH1,2), supplemented with a pad detector behind the RICHes.
In 1998, after removal of the pad detector, a major upgrade of the spectrometer 
was performed by the
addition of a large cylindrical Time Projection Chamber (TPC)
(see Fig.~\ref{Fig:setup}).
All subdetectors have a common acceptance in the 
polar angle range of $8^{\circ}<\theta<15^{\circ}$
at full azimuth,
corresponding to a pseudorapidity acceptance of $2.1<\eta<2.65$.

The SDD's are located about 12 cm downstream of the target system. 
Each SDD has a uniform radial drift field, with 360 equally spaced
readout anodes arranged along the outer perimeter of the Si-wafer.
A SDD track segment is defined by a hit in each of the two detectors
whose straight line connection points to the vertex of the interaction.

The TPC, located about 3.8~m downstream of the target, has an active 
length of 2~m and a diameter of 2.6~m. It is filled  
with a mixture of 80\% Neon and 20\% CO$_2$, and operated
inside the magnetic field generated by two opposite-polarity solenoidal coils,
which are placed around the TPC.
In the TPC drift volume, 
the ionization electrons created along charged particle tracks drift
outwards along the field lines of a radial electric drift field.
They are detected in 16 multi-wire proportional chambers with cathode pad
readout, which form the polygon-shaped outer termination of the active
TPC volume.
The TPC provides the reconstruction of up to 20 space points along
a curved charged particle track, thereby improving the momentum
resolution of the spectrometer and supplementing its particle 
identification capability via the measurement of the specific
energy loss d$E$/d$x$.

In 1999 (40~A~GeV/$c$ Pb-beam period) the target system consisted of eight
Au foils of 25 $\mu$m thickness, separated by 3.1~mm in beam direction 
and adding to 0.82\% of a hadronic interaction length.
In 2000 (80 and 158~A~GeV/$c$ Pb-beam period) the target was replaced
by 13 Au foils of the same dimensions and 2~mm spacing, with a total thickness
of 1.33\% of an interaction length.

The beam trigger (BEAM) is defined by the coincidence of two Beam
Counters BC1 and BC2 located in front of the target. 
Absence of a signal in a third Beam Counter BC3 located 
behind the target, in coincidence with BEAM, 
defines the interaction trigger (INT).
An online centrality selection was performed by an additional
threshold on the integrated pulse height in the SDD in 1999, or on the
signal in a scintillation Multiplicity Counter (MC) in 2000.
The data taking rate
was 300-500 events per burst, 
at a typical Pb beam intensity of $10^6$ ions per burst.
In 1999, the TPC readout was not yet fully operational. As a consequence,
the signals of only $60$\% of the 15360 TPC channels could be 
stored. 
The 80~A~GeV/$c$ data were taken while setting up for the
158~A~GeV/$c$ beam period, 
therefore no SDD information 
is available for the 80~A~GeV/$c$ data.

\section{Data analysis}
\subsection{Event selection and track reconstruction}
\label{Sec:event}

In the present analysis we have applied an offline centrality
selection of the upper 20\% of the total geometric cross 
section.
The centrality of the collision was defined by the number of charged particle
tracks reconstructed in the SDD (for the 40~A~GeV/$c$ data) or the 
pulse height in the Multiplicity Counter MC (for the 80 and 158~A~GeV/$c$ data).
The corresponding mean number of participating nucleons $\langle N_{\rm part} \rangle$
and mean number of nucleon-nucleon collisions $\langle N_{\rm coll} \rangle$ was
derived from a geometric nuclear overlap model using
$b=\sqrt{\sigma/ \pi}$ and resulting, with $\sigma_{NN}=30$~mb, 
in a total cross section of $\sigma_G=6.94$~b~\cite{eskola}.
Our classification  of central events comprises the 6.5\% most 
central fraction of the total geometric cross section.
For the centrality dependent studies we have subdivided our sample
into the 0-5\%, 5-10\%, 10-15\%, and 15-20\% most central
events (see Table~\ref{tab:cent-bins}).

 \begin{table}
 \centering
\vglue0.2cm
\caption{\label{tab:cent-bins} Definition of centrality classes.}
\vspace{0.25cm}
 \begin{tabular}{cccccc} \hline
$\sigma/\sigma_{\rm geo}$ & $0-6.5\%$ & $0-5\%$ & $5-10\%$ &  $10-15\%$ & $15-20\%$  \\ \hline
$\langle N_{\rm part} \rangle$ & 337 & 347 & 287 & 239 & 200 \\
$\langle N_{\rm coll} \rangle$ & 781 & 809 & 638 & 503 & 393 \\ \hline
\tabularnewline[5pt]
\end{tabular}
\end{table}

Additional offline cuts have been applied to reject pile-up events
and non-target interactions. The total number of analyzed events,
after centrality and quality cuts, is 
1.5M at 40~A~GeV/$c$ and 0.5M at 80 and 158~A~GeV/$c$, respectively.

SDD track information has only been used for the centrality definition
in 40~A~GeV/$c$ events, and for a number of systematic consistency
checks as described in Section~\ref{sec:sys}. All final results presented 
below are based on TPC track information only.

A TPC track is defined by a number of hits reconstructed
in subsequent readout-planes of the TPC and matching a 
momentum-dependent track model generated by a Monte Carlo simulation. 
Depending on the polar angle, a TPC track consists of up to
20 hits.
The momentum of the track is evaluated from the 
azimuthal deflection of the track inside the active TPC volume.
At the stage of the calibration used for this analysis, we achieved
a momentum resolution of about 
$\Delta p/p = ((1.5\%)^2 + (1.4\%\cdot p~({\rm GeV}/c))^2)^{1/2}$ 
for tracks containing 
more than 18 hits (in the 80 and 158~A~GeV/$c$ data sets).

A number of fiducial and quality cuts have been applied 
to provide stable tracking conditions and to reject tracks
from secondary particles: 
\begin{itemize}

\item	In the 40~A~GeV/$c$ data set, a fiducial cut in the 
	azimuthal angle  $17/24\pi < \phi < 2\pi$ is applied.
	Outside this region the efficiency was low due to 
   	malfunctioning read-out electronics during the 1999
	beam period. No such cut was applied for the 80 and 
	158~A~GeV/$c$ data from the year 2000, where
 	the read-out worked properly. The effect of the azimuthal
	fiducial cut on the final results has been studied 
	and is included in the systematic errors (see Section~\ref{sec:sys}).

\item   For the analysis of the mean $p_T$ fluctuations 
	only TPC tracks inside the full-length TPC track acceptance
	($2.2 < \eta < 2.7$) were used to provide sufficient $p_T$ resolution.
	For the same reason, the analysis was restricted to tracks with 
  	transverse momenta $0.1< p_T <1.5$~GeV/$c$.


\item 	The minimum number of hits per track is 11 at 40~A~GeV/$c$
	and 12 at 80 and 158~A~GeV/$c$ in the full-length TPC track acceptance
	$2.2 < \eta < 2.7$.

\item   To suppress secondary particles it is required that the back-extrapolation
 	of the particle trajectory into the target plane misses the interaction
	point by no more than 10~cm in transverse direction.

\item 	The $\chi^2/dof$ of the track momentum fit has to be less than 
	three times the r.m.s.~of the $\chi^2/dof$ distribution.
 
\end{itemize}

These cuts are used for all final results presented below. A systematic 
variation of these cuts has been applied to estimate our systematic 
uncertainties, as described in Section~\ref{sec:sys}.

\subsection{Measures of mean $p_{T}$ fluctuations}
\label{subsec:ptmeasure}
In the following, we briefly review quantities which have been
proposed as measures for event-wise mean $p_T$ fluctuations
and summarize notations used in this paper.

The event-wise mean $M_x$ of a single-particle observable $x$, 
averaged over particles $i$
in the acceptance of an event $j$, is given by: 

\begin{equation}
M_x^j \equiv \frac{\sum_{i=1}^{N_j}x_{i}}{N_j},
\end{equation}
where $N_j$ is the multiplicity of event $j$.

For a quantity $X_j$ which is defined for each 
event $j$ we calculate its mean over all events:

\begin{equation}
\langle X\rangle \equiv \frac{\sum_{j=1}^{n} w_{j} X_{j}}{n},
\end{equation}
where $n$ indicates the total number of events. 
The weighting factor $w_{j}$ for each event $j$ is 
equal to $N_{j}/\langle N \rangle$
for variables $X_j$ which are already
an average over $N_{j}$ particles, i.e.~$X_j = M_x^j$.
For other quantities $X_j$, such as multiplicity $N_j$,
the weighting factor $w_{j}$ is equal to 1.

The variance of the distribution of $X$ is given by:
\begin{equation}
\langle\Delta X^{2}\rangle \equiv \langle X^2 \rangle - \langle X\rangle^{2}.
\end{equation}
The inclusive mean (the mean over all particles in all events) and the
variance of a single particle observable $x$ are defined as:
\begin{equation}
\overline{x} \equiv \frac{\sum_{j=1}^{n}\sum_{i=1}^{N_{j}} x_{i}}{\sum_{j=1}^{n}N_{j}},
\end{equation}
and 
\begin{equation}
\overline{ \Delta x^2} \equiv \overline{x^2} - \overline{x}^2.
\end{equation}

The mean and the variance of $M_x$ are obtained by substituting 
$M_x$ for $X$ in the above equations, and including the event multiplicities
as appropriate weighting factors.
This weighting procedure provides
the most precise estimate of the variance of the parent
distribution in case of a finite mean multiplicity~\cite{Stephanov:1999}:
\begin{equation}
\langle M_x \rangle \equiv
\frac{\sum_{j=1}^{n}N_{j} M_x^j}
{\sum_{j=1}^{n}N_{j}}
= \overline{x}
\end{equation}
and 
\begin{equation}
\langle \Delta M_x^{2} \rangle \equiv \frac{\sum_{j=1}^{n}N_{j}(M_x^j-\langle M_x \rangle)^2
}{\sum_{j=1}^{n}N_{j}}.
\end{equation}


In the present event-by-event analysis, we search for dynamical
mean $p_{T}$ fluctuations beyond those expected in a purely statistical scenario.
Dynamical mean $p_{T}$ fluctuations would therefore result in 
an event-by-event distribution of $M_{p_{T}}$ which is wider
than that expected from the inclusive $p_{T}$ distribution
and the finite event multiplicity.

In previous analyses, the measure $\Phi_{p_{T}}$
has been used to quantify non-statistical mean $p_{T}$ fluctuations~\cite{Gazdzicki:1992}:
\begin{equation}
\Phi_{p_{T}} \equiv \sqrt{\frac{\langle Z^2 \rangle}{\langle N\rangle}}-\sqrt{\overline{z^2}},
\label{Eq:phipt-def}
\end{equation}
where $z$ and $Z$ are defined as
$z \equiv p_{T}-\overline{p_{T}}$ for each particle, and 
$\displaystyle Z \equiv \sum_{i=1}^{N} z_{i}$ for each event, respectively.
The measure $\Phi_{p_{T}}$ vanishes in the absence of correlations and dynamical 
mean $p_{T}$ fluctuations.
There is an approximate expression for $\Phi_{p_{T}}$ in terms of
the variances of the event-wise $M_{p_{T}}$ and the r.m.s. of the inclusive $p_T$ 
distributions~\cite{Voloshin:1999}:
\begin{equation}
\Phi_{p_{T}} \cong \sqrt{\langle N \rangle} \sqrt{\langle \Delta M_{p_{T}}^{2}\rangle}-\sqrt{\overline{\Delta p_{T}^{2}}}.
\end{equation}

A different measure for dynamical mean $p_T$ fluctuations has been
proposed in~\cite{Voloshin:1999}:
\begin{equation}
\sigma_{p_{T},dyn}^{2} \equiv \langle \Delta M_{p_{T}}^{2} \rangle -\frac{\overline{\Delta p_{T}^{2}}}{\langle N \rangle}.
\label{Eq:sdyn-def}
\end{equation}

This expression provides a direct relation between the variance
of the inclusive $p_T$ distribution, the mean multiplicity, and
the variance of the event-by-event mean $p_T$ distribution.
In case of vanishing non-statistical fluctuations and correlations, 
$\sigma_{p_{T},dyn}^{2}$ is equal to zero.
It has also been shown that $\sigma_{p_{T},dyn}^{2}$ is the mean of
the covariances
of all possible
pairs of two particles from the same event~\cite{Voloshin:1999}.

An alternative approach to evaluate $\sigma_{p_{T},dyn}^{2}$ is based
on the analysis of sub\-events~\cite{Voloshin:1999}. The
subevent method was applied for a number of consistency checks, however,
all final results presented below are obtained using Eq.~(\ref{Eq:sdyn-def})
for the evaluation of $\sigma_{p_{T},dyn}^{2}$.

There is an important relation between $\Phi_{p_{T}}$ and 
$\sigma_{p_{T},dyn}^{2}$~\cite{Voloshin:1999}:
\begin{equation}
\sigma_{p_{T},dyn}^{2} \cong
    \frac{2\Phi_{p_{T}}\sqrt{\overline{\Delta p_{T}^{2}}}}{\langle N \rangle}.
\label{Eq:phipt-sdyn}
\end{equation}

Based on our results presented below we find that this relation 
holds extremely well.

In order to account for a possible change of mean $p_{T}$ at 
different beam energies, we 
define a 
dimensionless measure, the "normalized dynamical fluctuation"~$\Sigma_{p_T}$:
\begin{equation} 
\Sigma_{p_T} \equiv 
{\rm sgn}({\sigma^{2}_{p_{T},dyn}})\cdot \frac{\sqrt{|\sigma^{2}_{p_{T},dyn}|}}{\overline{p_{T}}}.
\end{equation}

Due to the central limit theorem, 
the two measures $\Sigma_{p_T}$ and $\Phi_{p_{T}}$
are equal to zero in a purely statistical distribution.
If dynamical mean $p_T$ fluctuations are present, both measures
are finite and positive. However, also (anti-) correlations in momentum space
cause non-vanishing values. Long-range correlations occur as a consequence
of energy and momentum conservation, while Bose and Fermi statistics,
final state interactions, and experimental effects such as the
finite two-track resolution are the origin of short-range
correlations. Since such correlations mask the true fluctuation 
signal, it is crucial to investigate their contribution
to Eq.~(\ref{Eq:phipt-def}) and Eq.~(\ref{Eq:sdyn-def}) quantitatively, 
as will be described in Section~\ref{sect:pt-results}.

The best way for a quantitative determination of mean $p_T$ fluctuations
is clearly the use of a dimensionless measure.  
In addition, a comparison between different experiments should be 
possible. In this context we note that $\Phi_{p_{T}}$ is neither dimensionless
nor independent of the event multiplicity. This makes a comparison
between experiments and to theory difficult because the multiplicity
depends on the acceptance window of the experiment and 
on beam energy. Since
different contributions to the fluctuation signal have different
multiplicity dependences, the multiplicity dependence of $\Phi_{p_{T}}$
is {\it a priori} unknown. 

\begin{figure}[t]
\begin{center}
\mbox{\epsfig{file=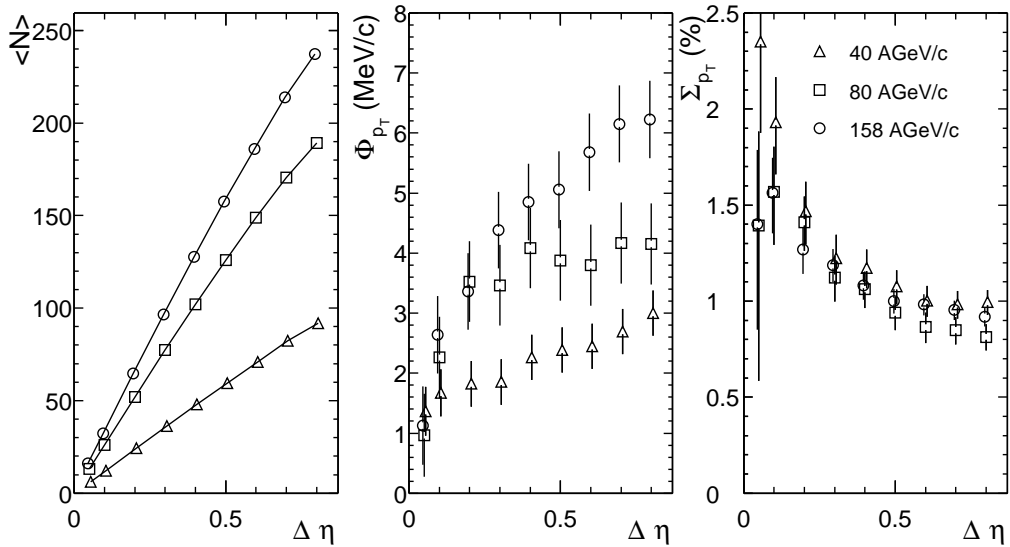,height=8cm}}
\end{center}
\caption{The mean charged particle multiplicity $\langle N \rangle$ (left)
and the fluctuation measures $\Phi_{p_{T}}$ (middle)
 and $\Sigma_{p_{T}}$
(right) in central Pb-Au events at 40, 80, and 158~A~GeV/$c$ 
as function of the pseudorapidity bin size $\Delta \eta$. 
The center of the $\Delta \eta$ window is always fixed at $\eta = 2.45$.
Note that the statistical errors are correlated.
}
\label{Fig:ebept-deta}
\end{figure}

In Fig.~\ref{Fig:ebept-deta} (left panel) the mean multiplicity is shown
as function of the size of the pseudorapidity
acceptance window $\Delta \eta$ at
40, 80, and 158~A~GeV/$c$. The much smaller multiplicity at 40~A~GeV/$c$
is related to the smaller azimuthal fiducial window in this data set.
The measure $\Phi_{p_{T}}$, shown in Fig.~\ref{Fig:ebept-deta} (middle panel),
exhibits a strong dependence on $\Delta \eta$ and beam energy. However,
comparison to the left panel in Fig.~\ref{Fig:ebept-deta} indicates
that this dependence is 
strongly correlated with
the change in
mean multiplicity.
In contrast, the normalized dynamical fluctuation
$\Sigma_{p_{T}}$ changes very little with beam energy, as 
shown in Fig.~\ref{Fig:ebept-deta} (right panel).
The increase of $\Sigma_{p_{T}}$ towards small $\Delta \eta$ 
may be attributed to 
physical correlations occuring on small $\Delta \eta$ scales.
These contributions appear to vanish at $\Delta \eta \geq 0.5$ 
and $\Sigma_{p_{T}}$ does not any longer 
depend on $\Delta \eta$, although the multiplicity
is still increasing\footnote{
We have checked that 
the definition of $\sigma_{p_{T},dyn}^{2}$
in terms of the mean of covariances
and the definition in Eq.~(\ref{Eq:sdyn-def})
lead to the same results for $\Sigma_{p_{T}}$ within statistical errors, 
also for small multiplicities. 
In particular,
the dependence on $\Delta \eta$ is observed using both definitions.
}.
This 
suggests that
$\Sigma_{p_{T}}$
is a more universal measure of fluctuations and it will be mainly used
in the present work. In Section~\ref{sect:pt-results} we will discuss
the implications of the results presented in this figure.


\subsection{Study of systematic uncertainties}
\label{sec:sys}
In this subsection, we briefly summarize the different contributions
to the systematic uncertainties of the fluctuation measurements
presented below.
The contributions to the systematic errors of the 
mean $p_T$ fluctuation measurements are
listed in Table~\ref{tab:pt-sys-err}.

Biases due to our analysis chain have been evaluated
with the help of the {\sc urqmd} (version 1.1) event generator ~\cite{UrQMD}
(without rescattering), which exhibits mean $p_T$ 
fluctuations
of similar magnitude as observed in the data. The generated
events were processed through a full Monte Carlo simulation
of the detector using the detector simulation package {\sc geant}~\cite{geant} 
including signal digitization,
and analyzed by our reconstruction software applying all cuts 
listed before. 
No significant bias on the initial fluctuation
signal was observed after event reconstruction.
This implies also that a small remaining contribution of secondary particles
does not alter the fluctuation signal.
\\
The systematic uncertainty due to the finite tracking
efficiency was evaluated using the same Monte Carlo chain
as before.
The number reported in Table~\ref{tab:pt-sys-err}
corresponds to the change of the reconstructed
$\Sigma_{p_{T}}$
when randomly
a fraction of the tracks (up to 20\%) is artificially 
removed from the sample. 
The estimated tracking efficiency in real data is 90\%.
\\
An online before- and after-protection of $\pm 1$~$\mu$s 
against a second beam particle was applied during data 
taking. In the offline analysis, we systematically increased
this protection to $\pm 3$~$\mu$s via event selection
but found very little effect on the reconstructed value
of $\Sigma_{p_{T}}$.
\\
The absolute momentum scale of the spectrometer has been verified
by the peak position of the invariant mass distributions of
reconstructed $\Lambda$ and $K_s^{\circ}$ decays~\cite{Tilsner:2002}.
At the present stage of the calibration, the uncertainty of the 
absolute momentum scale is at most $\pm 0.02$ $({\rm GeV/}c)^{-1}$ in $1/p$.
We have artificially shifted $1/p$ of each reconstructed
track by a constant offset to study the effect of this uncertainty.
As a test of the long-term stability of the momentum scale 
we divided our events into subsamples which were taken close
in time and analyzed them separately. After proper 
calibration the results of the subsamples were consistent 
with each other and with the result of the full data set.
\\
The impact of the azimuthal fiducial cut in the 40~A~GeV/$c$
data set
was studied by applying the same cut to the 158~A~GeV/$c$ data
and comparison to the full acceptance result.
We found that the contribution of the fiducial cut to
the final result is very small.
\\
The inclusion of SDD track information leads to a powerful rejection
of non-vertex tracks, if only TPC tracks with a match
to the SDD are used in the analysis. 
We compared the fluctuation results with and without
use of the SDD in the 158~A~GeV/$c$ data set. 
From the difference we estimated the systematic uncertainty
if SDD information is not used.
Since SDD information is not 
available for all data sets, we used TPC tracks
only for all final results presented below and included  
the uncertainty into the systematic error, assuming it is the
same at all beam energies.
\\
The systematic contribution of track quality cuts
was investigated by variation of these cuts ($\chi^2$, vertex) within
reasonable limits. The observed effect on the reconstructed value
of $\Sigma_{p_{T}}$ was also included into the systematic error.
\\
We developped a strategy to remove from our data set
contributions of short-range correlations (SRC) 
which may alter the measured fluctuation signal.
The occurence of short-range correlations
is due to quantum statistics, final state  
Coulomb interactions and the finite two-track resolution. 
The systematic uncertainty introduced by this procedure
is also listed in Table~\ref{tab:pt-sys-err}. For a detailed
description of the procedure see Section~\ref{sect:pt-results}.

The sum of all systematic uncertainties reported in 
Table~\ref{tab:pt-sys-err} is small compared to the experimentally observed
values of $\Sigma_{p_{T}} \approx 1\%$.

 \begin{table}
 \centering
\vglue0.2cm
\caption{\label{tab:pt-sys-err}Systematic errors of $\Sigma_{p_{T}}$
at $2.2 < \eta < 2.7$ in the 6.5 \% most central events.
}
\vspace{0.25cm}
 \begin{tabular}{cccc} \hline
       & 40 A~GeV/$c$& 80 A~GeV/$c$ & 158 A~GeV/$c$ \\ \hline
Tracking efficiency & $\pm 0.11$ \% & $\pm 0.11$ \% & $\pm 0.11$\%\\
Beam pile-up & $\pm 0.03$ \% & $\pm 0.03$ \% & $\pm 0.03$ \%\\
Absolute momentum scale      & ${}^{+0.08}_{-0.03}$ \% & ${}^{+0.05}_{-0.07}$ \% &
 ${}^{+0.02}_{-0.07}$ \% \\
Azimuthal fiducial cut & $\pm 0.02$ \% & -             & - \\
Non-vertex track contribution & $\pm 0.02$ \% & $\pm 0.02$ \% & $\pm 0.02$ \%  \\
$\chi^2$, vertex cut & ${}^{+0.39}_{-0.04}$ \% & ${}^{+0.13}_{-0.01}$ \% &
 ${}^{+0.10}_{-0.03}$ \%\\ \hline
SRC removal & ${}^{+0.07}_{-0.04}$ \% & ${}^{+0.05}_{-0.06}$ \% & ${}^{+0.02}_{-0.02} \%$\\ \hline
Total (no SRC removal) & ${}^{+0.42}_{-0.13}$ \% & ${}^{+0.18}_{-0.14}$
\% & ${}^{+0.15}_{-0.14}$ \%\\ \hline
Total (with SRC removal) & ${}^{+0.42}_{-0.14}$ \% & ${}^{+0.19}_{-0.15}$ \% & ${}^{+0.16}_{-0.14}$ \%\\ \hline
\tabularnewline[5pt]
\end{tabular}
\end{table}

\section{Results on mean $p_{T}$ fluctuations}
\label{sect:pt-results}

\begin{figure}[t]
\begin{center}
\mbox{\epsfig{file=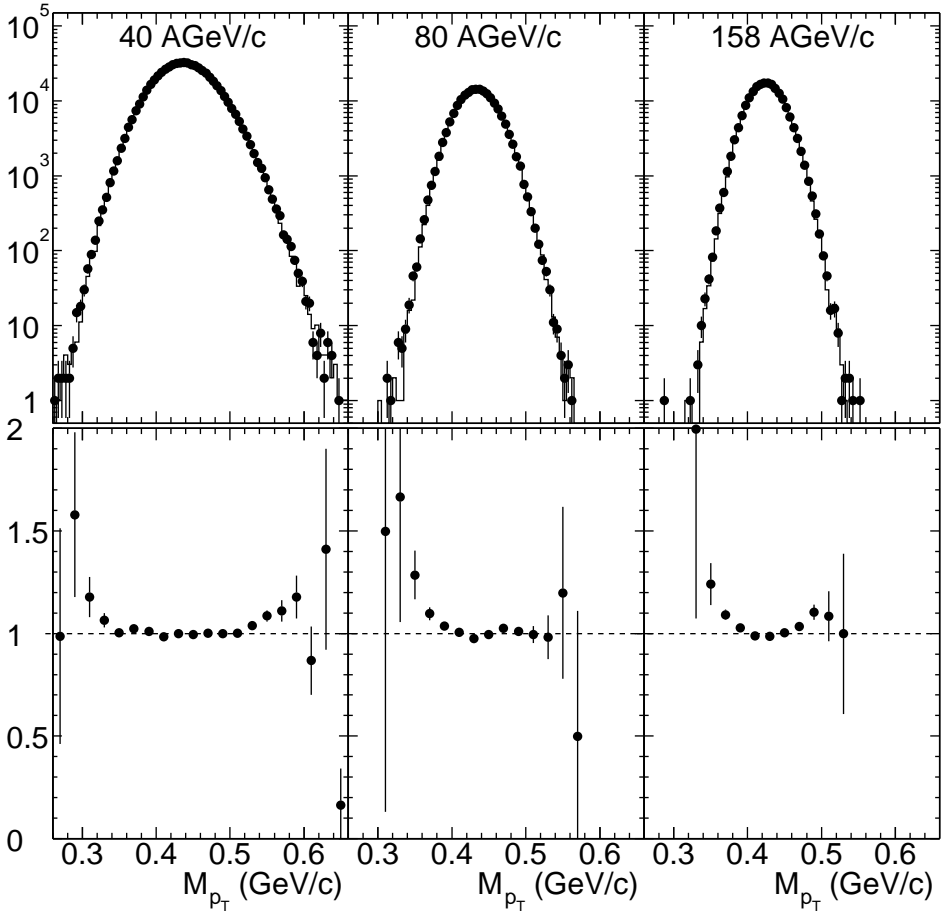,height=13cm}}
\end{center}
\caption{Top : Event-by-event mean $p_{T}$ distributions in the
6.5\% most central events at 40,
80, and 158~A~GeV/$c$. Circles show the distributions of data
events, solid lines indicate the mixed events. Bottom: Ratio between the distributions
of data events and mixed events for 40, 80, and 158~A~GeV/$c$.}
\label{Fig:ebept-mix}
\end{figure}

We obtain a first qualitative hint for the presence of non-statistical
mean $p_{T}$ fluctuations by an investigation of the event-by-event 
mean $p_{T}$ distributions as shown in Fig.~\ref{Fig:ebept-mix}. 
The larger width at 40~A~GeV/$c$ is a consequence of the smaller
azimuthal fiducial window and hence smaller multiplicity 
in this data set.
We compare the event-by-event mean $p_{T}$ distributions to reference
distributions obtained by event mixing. The mixed events are constructed
from particle momenta randomly chosen from data events of the same 
centrality class. Only one particle per measured event is used for a given
mixed event, and the multiplicity distribution of mixed events
is generated by sampling that of the data events.
We calculated $\Sigma_{p_{T}}$
and $\Phi_{p_{T}}$ for the mixed event samples and 
found them to be consistent with zero within statistical errors 
at all three beam energies.

The mixed event mean $p_{T}$ distributions exhibit a
Gamma distribution shape~\cite{Tannenbaum:2001}.
The subtle but clearly significant differences between the data and 
mixed event distributions
are emphasized in Fig.~\ref{Fig:ebept-mix}~(bottom), where the ratio
of the two is shown. The real event distributions are slightly wider,
indicating a small but finite non-statistical contribution to the  
mean $p_{T}$ fluctuations at all three energies. A preliminary
account of these results was presented in~\cite{Appelshauser:2002}.

Before we turn to a quantitative analysis 
of the observed non-statistical fluctuations, 
we discuss the effect of short-range correlations,
which are not present in our mixed event sample but
which may contribute to the fluctuation signal.
Physical short-range correlations are caused by Bose-Einstein
quantum statistics and the final state Coulomb interaction; 
detector effects such as the finite two-track resolution
lead to anti-correlations due to track merging.

\begin{figure}[t]
\begin{center}
\mbox{\epsfig{file=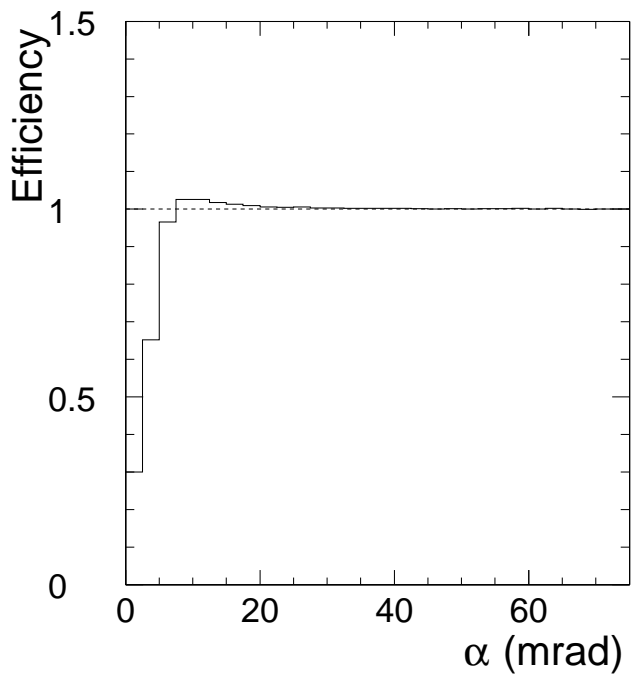,height=10cm}}
\end{center}
\caption{The two-track efficiency as function of the opening angle.}
\label{Fig:2track}
\end{figure}

In Fig.~\ref{Fig:2track} we present the opening angle distribution of
particle pairs detected in the TPC divided by the corresponding
mixed-event distribution. The resulting two-track detection
efficiency is normalized to unity at large opening angles $\alpha$.
The slight increase towards smaller $\alpha$ can be attributed to small
angle correlations such as Bose Einstein correlations and flow. 
At $\alpha<10$~mrad the efficiency drops due to track merging.

It was demonstrated by NA49~\cite{NA49:1999} that the effects of 
Bose-Einstein correlations and two-track resolution on the fluctuation signal
quantitatively cancel in their data set. In general, this is not 
necessarily the case but rather depends on the details
of the momentum range under investigation, the bending power of
the spectrometer and the two-track resolution.

Short-range correlations show up at small momentum differences
$q$ and can be investigated by a study of the two-particle
correlation function $C_2(q_{\rm inv})$. The four-momentum
difference 
$q_{\rm inv} \equiv
\sqrt{\mbox{\boldmath $q$}^2-q_{0}^2}$ is the momentum difference in
the pair rest frame, where \mbox{\boldmath $q$} and
$q_{0}$ are
the differences in three-momentum and energy
of a particle pair assuming the pion mass for each particle. 
A systematic analysis of the two-pion correlation
functions at SPS energies has been presented in~\cite{Adamova:2002}.

\begin{figure}[t]
\begin{center}
\mbox{\epsfig{file=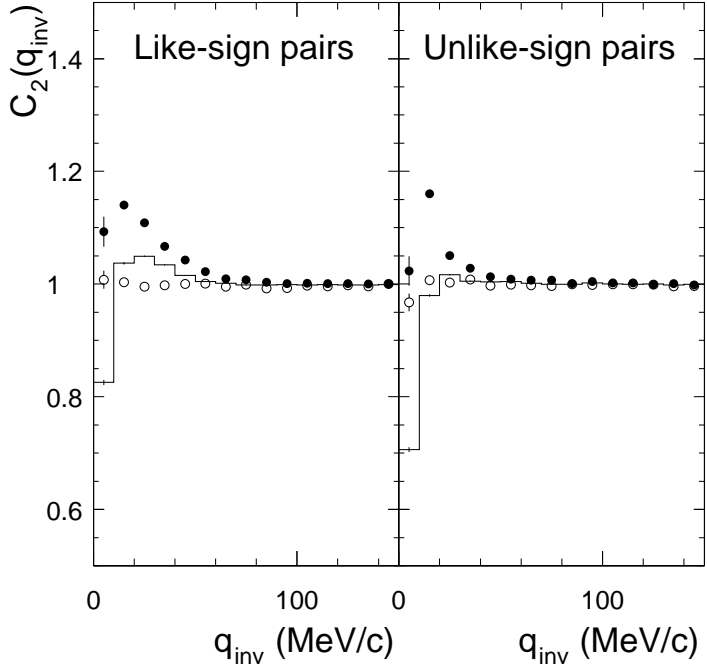,height=10cm}}
\end{center}
\caption{The correlation function $C_{2}(q_{\rm inv})$ 
as function of $q_{\rm inv}$ for like-sign pairs (left panel) 
and unlike-sign pairs
(right panel) at 158~A~GeV/$c$.
The histograms show raw data, 
the open circles are after removal of short-range correlations. 
The full dots show the reference correlation functions
with $\alpha > 10$~mrad (see text).}
\label{Fig:hbtcor80}
\end{figure}

The raw correlation functions $C_{2,{\rm raw}}(q_{\rm inv})$ for pairs of 
like-sign and unlike-sign charged particles in 158~A~GeV/$c$
are shown as histograms in Fig.~\ref{Fig:hbtcor80}. The like-sign pairs 
show a positive correlation at $q_{\rm inv}<70$~MeV/c due
to Bose-Einstein statistics in the pion-dominated sample.
At very small $q_{\rm inv}$, we observe an anti-correlation
caused by the final state Coulomb repulsion and the finite 
two-track resolution. For the unlike-sign pairs, the
positive correlation arises due to Coulomb
attraction, and is again suppressed by the two-track resolution
at small $q_{\rm inv}$. 
As a reference, we also constructed the correlation functions 
$C_2(q_{\rm inv})$ for signal and background pairs with
opening angles $\alpha > 10$~mrad (full dots in Fig.~\ref{Fig:hbtcor80}).
These correlation functions are free of distortions
from the finite two-track resolution but suffer from 
poor statistics at small $q_{\rm inv}$.

Based on the experimental two-track resolution and the 
measured correlation functions $C_2(q_{\rm inv})$, 
a procedure was applied to remove contributions
from short-range correlations on the observed fluctuation
signal.
The procedure implies a small modification of the measured
events. 
To account for the two-track resolution,
artificial tracks are added,
and  a small number of tracks from real pairs are removed
to suppress the positive correlation at small $q$. 
After the procedure, the correlation functions $C_2(q_{\rm inv})$
are flat, as indicated by the open symbols in
Fig.~\ref{Fig:hbtcor80}. From the modified events the fluctuation
signal is calculated, which is then expected to be free of contributions from 
short-range
correlations. The details of the procedure are described below.

In a first step 
tracks are randomly added to form, with existing tracks, 
small opening angle pairs 
according to the observed two-track reconstruction efficiency. 
The probability
to add a track at a given $\alpha$ is determined from the
distribution presented in Fig.~\ref{Fig:2track}
and constrained by the strength of the reference correlation 
function of pairs with $\alpha>10$~mrad in Fig.~\ref{Fig:hbtcor80}.
The fraction of artificially added tracks is a few percent, and
their transverse momenta are randomly chosen from tracks of different
events of the same centrality class.

In a second step, positive correlations are eliminated by 
removal of tracks from small $q_{\rm inv}$ pairs.
The probability to remove a track is based on
the measured reference correlation function $C_2(q_{\rm inv})$.
We define, 
for each pair of particles from the same event,
the probability to reject randomly one of the 
particles:

\[ p_{rej} =
b\left(1-\frac{1}{C_{2}(q_{\rm inv})}\right) \mbox{at } q_{\rm inv}<100~\mbox{MeV}/c, \]

where  $C_2(q_{\rm inv})$ is the reference correlation function of pairs
with $\alpha>10$~mrad and $b$ is a parameter tuned for like-sign
and unlike-sign pairs separately. After both steps, the obtained
correlation functions are flat, as indicated by the open circles in Fig.~\ref{Fig:hbtcor80},
for typical values of $b$ around $0.8$-$1.0$. 

\begin{figure}[t]
\begin{center}
\mbox{\epsfig{file=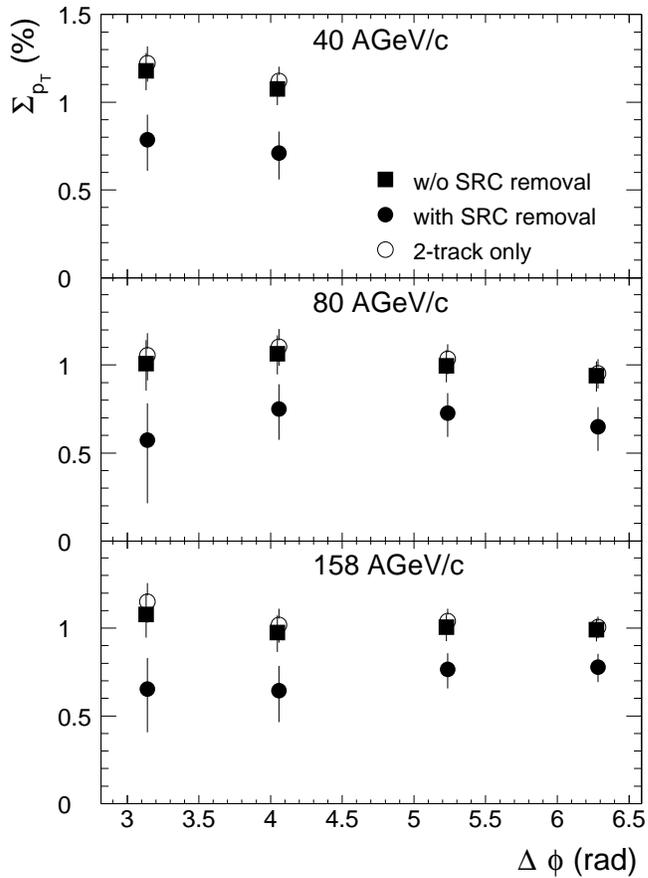,height=12cm}}
\end{center}
\caption{The fluctuation measure $\Sigma_{p_{T}}$
as function of the azimuthal acceptance window $\Delta \phi$ in central
Pb-Au collisions at 40, 80,
and 158~A~GeV/$c$.
The filled squares are without SRC removal, the filled circles are after
the full SRC removal procedure.
The open circles show the results obtained after compensation of
the two-track resolution only.  }
\label{Fig:ebept-hbtcor}
\end{figure}

In Fig.~\ref{Fig:ebept-hbtcor} is presented the normalized dynamical fluctuation
$\Sigma_{p_{T}}$
in central Pb-Au events at 40, 80,
and 158~A~GeV/$c$ as function of the size of the azimuthal acceptance
window $\Delta \phi$. 
The filled squares show the result for the unmodified events.
If additional tracks are added to compensate for the two-track
resolution, the results increase very slightly, as indicated by the
open symbols. 
The filled dots show the results after the full SRC removal,
which contains the compensation
for the two-track resolution and the suppression of correlated small $q$
pairs. The contributions from both effects do not cancel but 
lead to a reduction of $\Sigma_{p_{T}}$ by $20-35$\% compared to the values
obtained from unmodified events. 
The systematic errors
of the correction procedure have been estimated to be less than 0.05\% (absolute)
in most cases
by variation of the probabilities to add or reject a track within reasonable
limits. 
They are small compared to the overall systematic uncertainties
discussed before (see Table~\ref{tab:pt-sys-err}).

Note that the azimuthal acceptance of the 40~A~GeV/$c$ data set is limited,
but no significant change of the fluctuation strength is observed when 
going from limited to full acceptance at the higher beam energies. 
This observation is the basis of the estimate of the systematic error
introduced by the limited acceptance at 40~A~GeV/$c$ as discussed
in Section~\ref{sec:sys}. This implies also that 
azimuthal anisotropies of the particle distribution, such as
elliptic flow, have little effect on the observed 
fluctuation strength. 

As a consistency check of the SRC removal procedure 
we applied the subevent method~\cite{Voloshin:1999} to calculate
the fluctuation $\Sigma_{p_{T}}$.
If the subevents are based on a random selection of tracks from
a given event, the observed fluctuation $\Sigma_{p_{T}}$ is consistent
with the result obtained by our standard method using Eq.~(\ref{Eq:sdyn-def})
without SRC removal. 
The contribution from short range correlations can be estimated
if the subevents are separated in pseudorapidity.
We have chosen the pseudorapidity ranges $2.2<\eta<2.4$ and
$2.5<\eta<2.7$ for subevents 1 and 2, respectively.
In this case, the observed mean $p_T$ fluctuations are 
reduced compared to the case of random subevents, and consistent 
with those obtained using our previously described method for 
SRC removal.

To facilitate a comparison of the data to models, which generally
do not contain short-range correlations, 
only the results after SRC removal are shown below.
A compilation of all results before and after SRC removal can be found in 
Table~\ref{tab:ptsumc6.5} and in Appendix~\ref{App:tab-pt}.

In Fig.~\ref{Fig:ebept-cent} the normalized dynamical 
fluctuation $\Sigma_{p_{T}}$ is shown
as function of the centrality of the collision at three different 
beam energies. 
The centrality is expressed in terms of the average number of participating
nucleons $\langle N_{\rm part} \rangle$ in a given centrality bin. 
We observe that, at all energies,
the dynamical fluctuation $\Sigma_{p_{T}}$ is 
comparable in strength and decreases with centrality
by approximately 40\% from about 1.3\% to 0.7\% over the 
centrality range under investigation.

\begin{figure}[t]
\begin{center}
\mbox{\epsfig{file=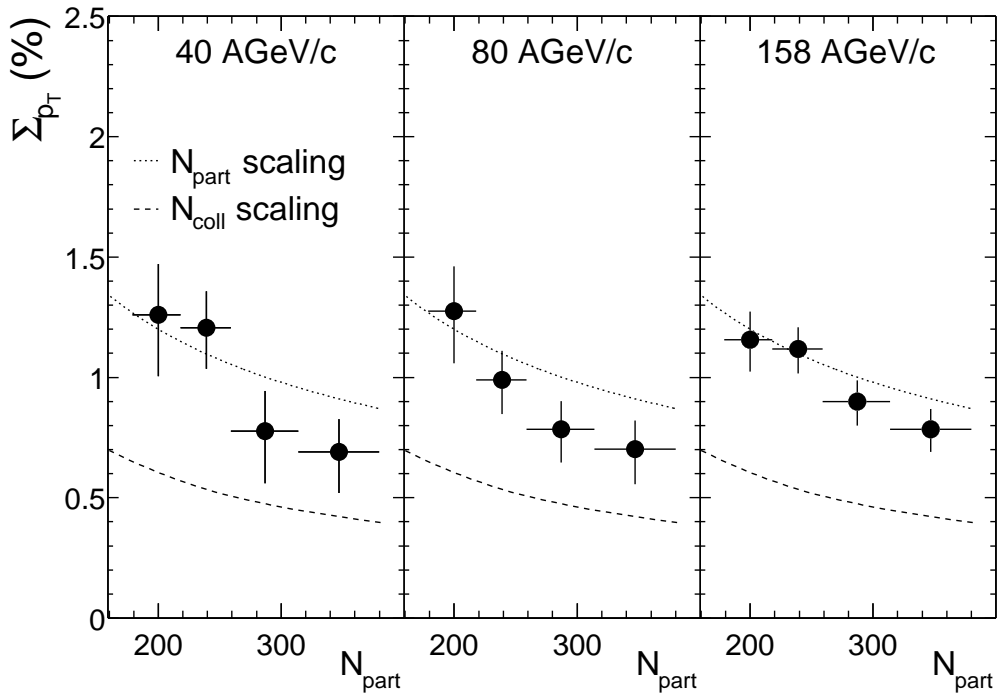,height=10cm}}
\end{center}
\caption{The fluctuation measure $\Sigma_{p_{T}}$
as 
function of the number of participants in Pb-Au collisions at 40, 80,
and 158~A~GeV/$c$. 
The data points are obtained after removal of short-range correlations.
Also shown are extrapolations from $pp$ data assuming $N_{\rm part}$ scaling
(dotted lines) and $N_{\rm coll}$ scaling (dashed lines, see text).}
\label{Fig:ebept-cent}
\end{figure}

In this context, it is interesting to compare the present results to 
measurements in hadron-hadron collisions. In $pp$ interactions particles are
produced in a correlated way which leads to large non-statistical
fluctuations. At the ISR, dynamical mean $p_t$ fluctuations have been measured
in $pp$ reactions at $\sqrt{s_{NN}}=30.8-63.0$~GeV~\cite{braune_pp}.
Independent of beam energy, a value of 12\% was observed for 
$\Sigma_{p_{T}}$.
In $\alpha$$\alpha$ reactions,
the observed dynamical fluctuation is reduced to about 9\%. This was 
stated in~\cite{braune_pp} 
to be consistent with the larger average number of nucleon-nucleon
interactions in $\alpha$$\alpha$ of about 1.8. 
Under the assumption of an incoherent
superposition of independent nucleon-nucleon collisions, 
$\Sigma_{p_{T}}$ is expected to scale with
$\langle N_{\rm coll} \rangle$:

\begin{equation}
  \Sigma_{p_{T}}^{\rm AA} = \Sigma_{p_{T}}^{\rm pp}
        \cdot \langle N_{\rm coll} \rangle ^{-1/2}.
\label{eq:ncollscale}
\end{equation}

Use of this expression to extrapolate from the dynamical fluctuation of 
12\% observed in $pp$ ($N_{\rm coll}=1$) 
to Pb-Au yields values of about 0.4\% for the most central events.
These extrapolated numbers are significantly below the data, as shown in
Fig.~\ref{Fig:ebept-cent},
indicating that
A-A collisions at SPS energies are not a straight superposition of nucleon-nucleon
collisions also in this observable.

A different approach is a possible scaling with charged particle multiplicity:

\begin{equation}  
	\Sigma_{p_{T}}^{\rm AA} = \Sigma_{p_{T}}^{\rm pp}
        \cdot \left (\frac{\langle N_{\rm pp}\rangle}{\langle N_{\rm AA}\rangle}\right )^{1/2}.
\end{equation}

Since the number of charged particles was found to scale close to
linear with the number of participants $\langle N_{\rm part} \rangle$ at 
SPS~\cite{WA98,miskomult,WA97},
we replace $(\langle N_{\rm pp}\rangle/\langle N_{\rm AA}\rangle)^{1/2}$
by $(\langle N_{\rm part} \rangle /2)^{-1/2}$:

\begin{equation}
  \Sigma_{p_{T}}^{\rm AA} = \Sigma_{p_{T}}^{\rm pp}
        \cdot (\langle N_{\rm part} \rangle/2) ^{-1/2}.
\label{eq:npartscale}
\end{equation}

A comparison of Eq.~(\ref{eq:npartscale})
to our data is indicated by the dotted curve in Fig.~\ref{Fig:ebept-cent}. 
Although our data are at still somewhat smaller $\sqrt{s_{NN}}$ than
explored in~\cite{braune_pp},
we find that the magnitude of the fluctuation strength observed
in Pb-Au is similar to the extrapolation from $pp$, if multiplicity
scaling is assumed. 

\begin{figure}[t]
\begin{center}
\mbox{\epsfig{file=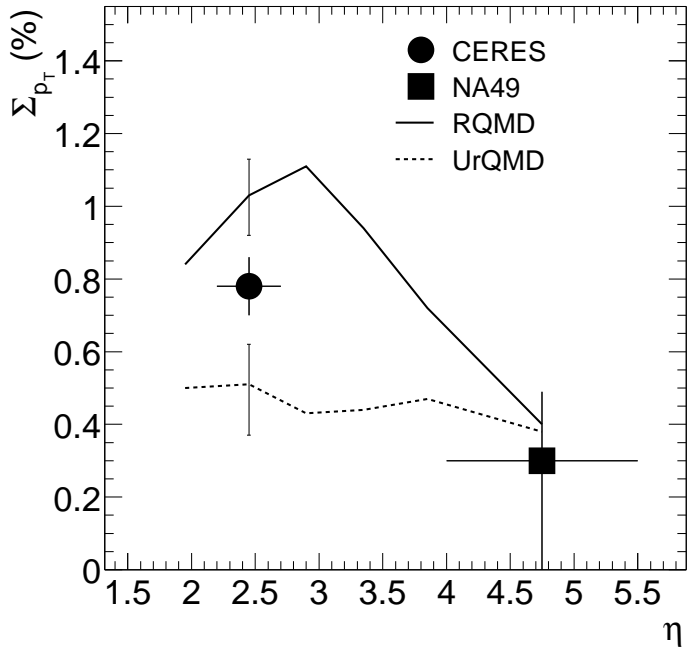,height=10cm}}
\end{center}
\caption{The fluctuation measure $\Sigma_{p_{T}}$
as function of pseudorapidity in central Pb-Au collisions at
158~A~GeV/$c$. 
The CERES data are after SRC removal.
Results from {\sc rqmd} and
{\sc urqmd} calculations are shown as solid and dashed
lines, respectively. The statistical errors of the calculations
are indicated by the error bars at $\eta=2.45$.
The NA49 data point for the 5\% most central Pb-Pb collisions 
($0.005<p_{T}<1.5$~GeV/$c$)
is calculated from~\cite{NA49:1999} (see text). 
The horizontal error bars indicate the $\eta$ acceptance range for 
the data points.
Note that NA49 data were measured in the pion rapidity interval $4<y_{\pi}<5.5$.}
\label{Fig:rqmd-phipt}
\end{figure}

In Fig.~\ref{Fig:rqmd-phipt} we compare our mid-rapidity result for 
$\Sigma_{p_{T}}$
in central Pb-Au collisions at 158~A~GeV/$c$ to the measurement 
from NA49 obtained at forward rapidity~\cite{NA49:1999}. 
The NA49 result for $\Phi_{p_T}$~\cite{NA49:1999} was used to evaluate
$\Sigma_{p_{T}}$ using Eq.~(\ref{Eq:phipt-sdyn}). 
While the mean $p_T$ fluctuations measured at forward rapidity are consistent
with zero, we observe a finite value of about 0.8\% at mid-rapidity.
Also shown are calculations
from the {\sc rqmd}~\cite{RQMD} and {\sc urqmd}~\cite{UrQMD} event generators.
While the fluctuations observed
in {\sc urqmd} are small at all rapidities, {\sc rqmd}
exhibits a pronounced rapidity dependence which qualitatively
describes the trend observed in the data.
However, the {\sc rqmd} absolute value around mid-rapidity is about 20\% larger than that
observed in the data.

\begin{figure}[t]
\begin{center}
\mbox{\epsfig{file=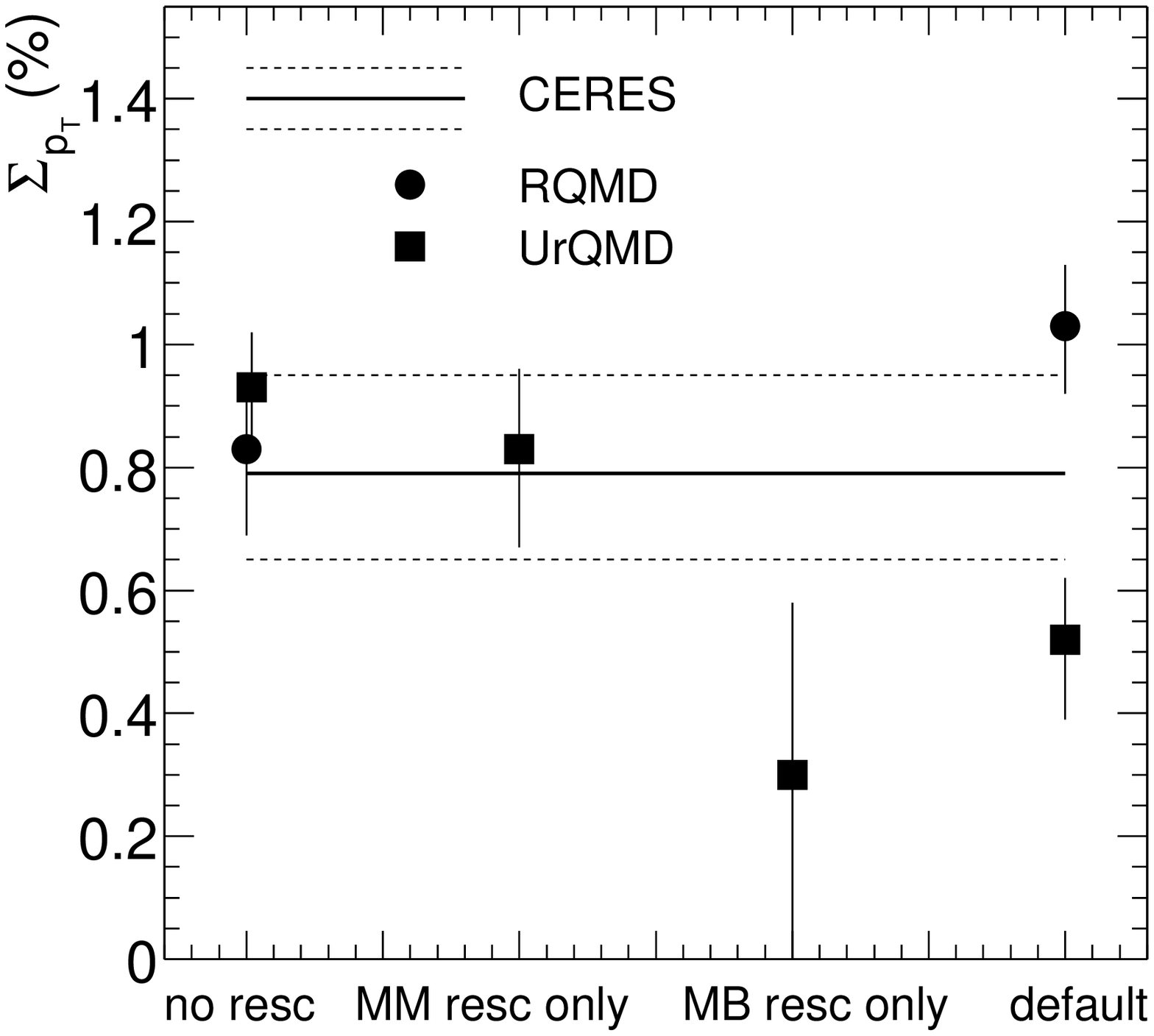,height=10cm}}
\end{center}
\caption{Model calculations of $\Sigma_{p_{T}}$
from {\sc urqmd} and {\sc rqmd} in comparison
to the CERES measurement (after SRC removal) at $2.2<\eta<2.7$ in central
Pb-Au collisions at 158~A~GeV (see text). 
The results of the model calculations are shown for different rescattering 
scenarios.
The dashed lines indicate 
the systematic error of the data.}
\label{Fig:mod_syst}
\end{figure}

The magnitude of mean transverse momentum fluctuations
from {\sc rqmd} and {\sc urqmd} was studied in more detail, as shown in
Fig.~\ref{Fig:mod_syst}. 
We compare our result for $\Sigma_{p_T}$ in central
158~A~GeV/$c$ Pb-Au collisions to calculations from both
models at $2.2 < \eta < 2.7$. 
Different mechanisms included in the models have been
investigated separately. 
If hadronic rescattering is 
switched off, both models show reasonable agreement with
the data. 
In {\sc rqmd}, the inclusion of rescattering leads to an increase
of fluctuations.
This behaviour is not understood and needs further investigation.
In {\sc urqmd}, the inclusion of meson-meson rescattering
has very little effect on the observed fluctuations, however,
inclusion of meson-baryon rescattering reduces the fluctuations
considerably. If both rescattering modes are included in {\sc urqmd}
(labelled 'default' in the figure), {\sc urqmd} falls below the 
data. 
We checked that also at 40 and 80~A~GeV/$c$ the agreement 
between data and {\sc urqmd} without rescattering is reasonable,
while inclusion of rescattering in {\sc urqmd} leads to 
a significant reduction of $\Sigma_{p_T}$. 
At 40~A~GeV/$c$, rescattering leads even to
negative values for $\Sigma_{p_T}$.
This indicates that secondary meson-baryon scattering 
has a strong effect on the observed fluctuation strength, 
and might be over-estimated in {\sc urqmd}.
It should be emphasized that the evaluation of event-by-event
fluctuations may provide important information about
the dynamics of the system which can not be derived
from inclusive observables.

\begin{figure}[t]
\begin{center}
\mbox{\epsfig{file=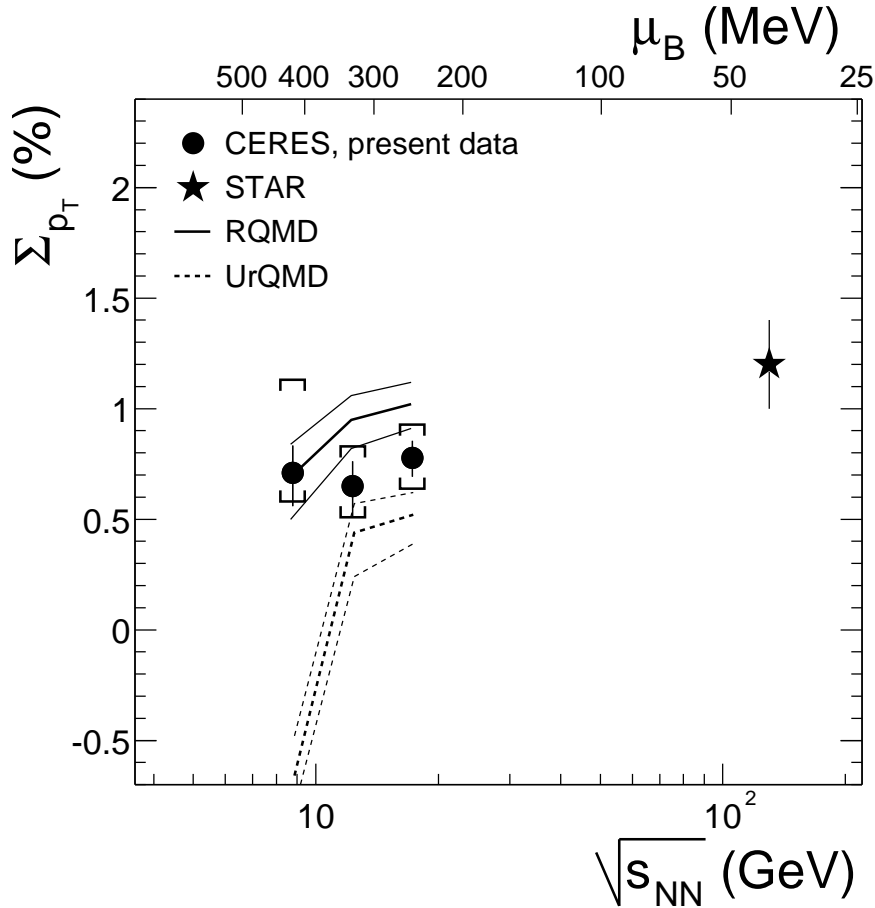,height=12.5cm}}
\end{center}
\caption{The fluctuation measure
$\Sigma_{p_{T}}$
as function of $\sqrt{s_{NN}}$ and of $\mu_B$ at chemical 
freeze-out~\cite{pbm-cley}.
The full circles show CERES results (after SRC removal) in central events
at 40, 80, and 158~A~GeV/$c$.
The brackets indicate the systematic errors.
Also shown is the STAR result~\cite{Voloshin:2001} at $\sqrt{s_{NN}}$ = 130~GeV
which is not corrected for SRC.
Results and statistical errors from {\sc rqmd} and
{\sc urqmd} calculations (with rescattering) are indicated as solid and dashed
lines, respectively.
}
\label{Fig:ebept-s}
\end{figure}

A compilation of the normalized dynamical fluctuation 
$\Sigma_{p_{T}}$ measured at mid-rapidity and at different 
beam energies is shown in Fig.~\ref{Fig:ebept-s}. 
The upper scale indicates the baryon chemical potential $\mu_B$
at chemical freeze-out, related to $\sqrt{s_{NN}}$
via a phenomenological parametrization given in~\cite{pbm-cley}.
The data point
shown at $\sqrt{s_{NN}}=130$~GeV is from the STAR
experiment at RHIC~\cite{Voloshin:2001} and measured over a somewhat wider 
$p_T$ range ($0.1<p_T<2$~GeV/$c$). 
A measurement by PHENIX~\cite{PHENIX-pt:2002} yields
$\Sigma_{p_{T}}=1.4^{+0.6}_{-1.8}$\%
and is consistent within its larger error.\footnote{The measure
$\Sigma_{p_{T}}$ was calculated using the PHENIX measurement for $d$ and $\omega_{p_T}$
from~\cite{PHENIX-pt:2002} and the approximation
$\Sigma_{p_{T}}^{2} \simeq 2 d \omega_{p_{T}}$.}
At the SPS, the normalized dynamical fluctuation $\Sigma_{p_{T}}$
is about 0.7\%, while larger fluctuations of 1.2\% are observed
at RHIC.
The STAR data are not corrected for the contribution of
short-range correlations, however, the  effect was estimated 
to be small (less than 10\%)~\cite{Voloshin:2001}.
The evolution of $\Sigma_{p_{T}}$
with beam energy looks smooth and does not show any indication
of unusually large fluctations at any beam energy. 

Models predict enhanced mean $p_T$ fluctuations if the system
has passed close to the critical point of the QCD phase diagram.
At SPS energies and for the finite rapidity acceptance
window of the CERES experiment, the fluctuations should
reach values of about 2\%,
i.e.~more than three times larger
than observed in the present data\footnote{
The predicted fluctuations in the measure $\sqrt{F} = 1.1$ in~\cite{Stephanov:1999} 
corresponds to about 2\% in $\Sigma_{p_{T}}$ in the CERES 
acceptance~\cite{Stephanov:2001zj}.
}. 
Most important, no
indication for a non-monotonic behaviour as function
of the beam energy has been observed. This suggests
that the critical point may not be located in the $\mu_B$
regime below 450~MeV.

The results from {\sc rqmd} and {\sc urqmd} show rough agreement
with the data, except for the {\sc urqmd} calculation at 40~A~GeV/$c$
where $\Sigma_{p_{T}}$ is negative (see Fig.~\ref{Fig:ebept-s}). 
We note that a positive value of $\Sigma_{p_T}=0.38^{+0.17}_{-0.48}\%$
is obtained from {\sc urqmd} at 40~A~GeV/$c$, if rescattering is switched off.

 \begin{table}
 \centering
\tabcolsep=1pt
\begin{tabular}{cccc} \hline
Beam momentum (A~GeV/$c$) & $40$ & $80$ & $158$ \\ \hline
$\Sigma_{p_{T}}$ (\%) & & & \\
(no SRC removal) & $1.08\pm 0.09^{+0.42}_{-0.13}$ &
$0.94^{+0.08}_{-0.09} {}^{+0.18}_{-0.14}$ &
$0.99^{+0.06}_{-0.07} {}^{+0.15}_{-0.14}$ \\
(with SRC removal) &
$0.71^{+0.12}_{-0.15} {}^{+0.42}_{-0.14}$ &
$0.65^{+0.11}_{-0.14} {}^{+0.19}_{-0.15}$ &
$0.78 \pm 0.08 {}^{+0.16}_{-0.14}$ \\ \hline
$\Phi_{p_{T}}$ (MeV/$c$) & & & \\
(no SRC removal) & $2.4 \pm 0.4^{+1.5}_{-1.2}$ &
$3.9 \pm 0.7^{+1.0}_{-0.6}$ &
$5.0 \pm 0.6^{+2.1}_{-1.2}$ \\
(with SRC removal) &
$1.1 \pm 0.4^{+1.5}_{-1.2}$ &
$1.9 \pm 0.7^{+1.1}_{-0.7}$ &
$3.1 \pm 0.6^{+2.1}_{-1.2}$ \\ \hline
$n$ & 618958 & 186272 & 199304 \\
$\langle N \rangle$ & $59.47 \pm 0.01$ & $126.01 \pm 0.04$ & $157.67 \pm 0.05$ \\
$\overline{p_{T}}$ (MeV/$c$) & $438.67 \pm 0.05$ & $434.68 \pm 0.06$ & $425.20 \pm 0.05$  \\
$\sqrt{\overline{\Delta p_{T}^{2}}}$ (MeV/$c$) & $290.14 \pm 0.05$ & $284.17 \pm 0.06$ & $278.97 \pm 0.05$ \\
$\sqrt{\langle \Delta M_{p_{T}}^{2} \rangle}$ (MeV/$c$) & $37.919$ $\pm 0.006$ & $25.642$ $\pm 0.005$ & $22.613$ $\pm 0.004$  \\ \hline
\tabularnewline[5pt]
\end{tabular}
\caption{Summary of mean $p_{T}$ fluctuations 
in the 6.5~\% most central Pb-Au events at 40, 80, and 158~A~GeV/$c$.
Statistical and systematic errors are quoted for $\Sigma_{p_T}$ and
$\Phi_{p_{T}}$. The other quantities are without SRC removal and 
statistical errors only are shown.
}
\label{tab:ptsumc6.5}
\vspace{1cm}
\end{table}

Mean transverse momentum fluctuations
can be related to event-by-event fluctuations of the temperature. The following expression
was given in~\cite{korus}:
\begin{equation}
	\Phi_{p_{T}} = \sqrt{2}\,\,\langle N \rangle \,\, \frac{\sigma_T^2}{\langle T \rangle},
\end{equation}
where $\langle T \rangle$ is the average temperature which fluctuates event by event 
with a standard deviation $\sigma_T$. 
To derive an upper limit for the temperature fluctuations, it is assumed 
that the full magnitude
of the observed mean $p_T$ fluctuations after removal of short-range correlations
is due to temperature fluctuations.
For a temperature $T=160$~MeV,
values for $\sigma_T$ of 1.4$\pm$0.3~MeV, 1.3$\pm$0.3~MeV, and 1.5$\pm$0.2~MeV 
are obtained in the 6.5\% most central events
at 40, 80, and 158~A~GeV/$c$, respectively. For $T=120$~MeV, the results are
smaller by about 15\%.
We conclude that event-by-event temperature fluctuations are at most
1\% of the average temperature. Such small numbers point to a large degree of
thermalization of the system. We note that small temperature
fluctuations are expected also in the case of complete thermalization~\cite{Gazdzicki:1999,korus}
and may in that case be used to determine the heat capacity of the system~\cite{korus}.

\section{Conclusions}

We have presented event-by-event fluctuations of the mean transverse momentum 
in Pb-Au collisions at 40, 80, and 158~A~GeV/$c$. 
In central events at these beam energies, significant non-statistical
positive mean $p_{T}$ fluctuations are observed.
Based on the measured two-particle correlation functions and
the experimental two-track resolution a procedure was derived
which accounts for the contributions of short-range correlations present in the data.
This procedure reduces the observed mean $p_T$ fluctuations by
20-35\% but they remain significantly different from zero.
At all beam energies, we find for the normalized dynamical
fluctuation $\Sigma_{p_{T}}$ values of about 0.7\% of the mean $p_T$. 
These results at mid-rapidity are complementary to previous measurements
at 158~A~GeV/$c$ in the forward hemisphere ($4<y_{\pi}<5.5$) where
no significant non-statistical fluctuations were found~\cite{NA49:1999}. 

The present mid-rapidity results at SPS 
are somewhat smaller than the values measured by STAR at 
$\sqrt{s_{NN}} = 130$~GeV.
The existing data from different beam energies do not exhibit an
indication for unusually large fluctuations or a non-monotonic 
behaviour, which might have pointed 
to the crossing of the critical point of the QCD phase diagram somewhere 
in this range of beam energy or baryon chemical potential~\cite{Stephanov:1999}. 
Further studies at lower beam energies are needed for a full exploration
of the QCD phase diagram.

At all three beam energies under investigation we observe a systematic
decrease of the mean $p_{T}$ fluctuation strength with increasing
centrality of the collision. 
Based on measurements of non-statistical mean $p_{T}$ fluctuations
in $pp$-collisions at the ISR we find that the centrality dependence 
of mean $p_{T}$ fluctuations
in Pb-Au is consistent with an extrapolation
from $pp$, assuming that mean $p_{T}$ fluctuations scale with
$\langle N \rangle^{-1/2}$.
This may suggest that the fluctuation pattern is not reduced strongly
by potential rescattering of hadrons after hadronization.
A comparison to results from the {\sc rqmd} and {\sc urqmd} models
indicates that secondary rescattering, if enabled, tends to decrease the
fluctuation strength, while calculations without rescattering
show reasonable agreement with the data.
Such a scenario is supported by the observation of high 
densities and a short mean free path at thermal freeze-out,
as derived from a recent analysis of pion 
interferometry data~\cite{Adamova:2003},
which point to a short lifetime of the hadronic phase. 
This issue needs further investigations of smaller
collision systems and minimum bias Pb-Au data, as well as
more detailed model studies.


\section{Acknowledgements}
The CERES collaboration acknowledges the good performance of the CERN
PS and SPS accelerators as well as the support from the EST division.
We are grateful for excellent support for the central data recording
from the CERN IT division. 
We wish to thank K.~Rajagopal, M.A.~Stephanov, and S.~Voloshin 
for valuable discussions.
This work was supported by the German BMBF,
the U.S.~DoE, the Israeli Science Foundation, and the MINERVA Foundation.

\newpage

\appendix

\section{Summary of mean $p_{T}$ fluctuations}
\label{App:tab-pt}
In this appendix the results of mean $p_{T}$
fluctuations in Pb-Au collisions at 40, 80, and 158~A~GeV/$c$ 
and different centrality classes are summarized. 
The results for $\Sigma_{p_{T}}$
and $\Phi_{p_{T}}$ are listed before
and after removal of short range correlations (SRC).
They contain statistical and systematic errors.
All other parameters are before SRC removal and statistical
errors only are shown.

 \begin{table}[h!]
 \centering
\vglue0.2cm
\caption{Summary of mean $p_{T}$ fluctuations at
40~A~GeV/$c$.
}
\vspace{0.25cm}
\tabcolsep=1pt
\begin{tabular}{ccccc} \hline
Centrality & $0-5$ \% & $5-10$ \% & $10-15$ \% & $15-20$ \% \\ \hline
$\Sigma_{p_{T}}$(\%) & & & & \\
(no SRC removal) &
$1.04^{+0.10}_{-0.11} {}^{+0.51}_{-0.26}$&
$1.18^{+0.12}_{-0.13} {}^{+0.33}_{-0.22}$&
$1.54^{+0.12}_{-0.13} {}^{+0.31}_{-0.16}$&
$1.68^{+0.16}_{-0.18} {}^{+0.32}_{-0.25}$\\
(with SRC removal)
&$0.69 {}^{+0.14}_{-0.17} {}^{+0.52}_{-0.27}$
&$0.78 {}^{+0.17}_{-0.21} {}^{+0.34}_{-0.23}$
&$1.21 {}^{+0.15}_{-0.17} {}^{+0.31}_{-0.16}$
&$1.26 {}^{+0.21}_{-0.26} {}^{+0.32}_{-0.25}$\\ \hline
$\Phi_{p_{T}}$ (MeV/$c$) & & & & \\
(no SRC removal)
&$2.3 \pm 0.4^{+2.3}_{-0.5}$& $2.4 \pm 0.5^{+1.4}_{-0.5}$
&$3.3 \pm 0.5^{+1.4}_{-0.4}$& $3.3 \pm 0.6^{+1.8}_{-0.7}$\\
(with SRC removal)
&$1.1 \pm 0.4^{+2.3}_{-0.5}$&$1.1 \pm 0.5^{+1.4}_{-0.5}$
&$2.0 \pm 0.5^{+1.4}_{-0.4}$&$1.9 \pm 0.6^{+1.8}_{-0.7}$\\ \hline
$n$
& 495951 & 396868 & 327473 & 214202\\
$\langle N \rangle$ &
$61.19 \pm 0.01$ & $49.39 \pm 0.01$ &
$40.94 \pm 0.01$ & $33.88 \pm 0.01$\\ 
$\overline{p_{T}}$ (MeV/$c$) &
$439.03 \pm 0.05$ & $436.07 \pm 0.07$ &
$433.15 \pm 0.08$ & $429.75 \pm 0.11$\\ 
$\sqrt{\overline{\Delta p_{T}^{2}}}$ (MeV/$c$) &
$290.35 \pm 0.05$ & $288.71 \pm 0.07$ &
$286.90 \pm 0.08$ & $284.84 \pm 0.11$\\ 
$\sqrt{\langle \Delta M_{p_{T}}^{2} \rangle}$ (MeV/$c$) &
$37.395 \pm 0.007$ & $41.400 \pm 0.009$ &
$45.333 \pm 0.012$ & $49.464 \pm 0.018$\\\hline
\tabularnewline[5pt]
\end{tabular}
\label{tab:ptsum40}
\end{table}

 \begin{table}[h!]
 \centering
\vglue0.2cm
\caption{Summary of mean $p_{T}$ fluctuations at 80~A~GeV/$c$.
}
\vspace{0.25cm}
\tabcolsep=1pt
\begin{tabular}{ccccc} \hline
Centrality & $0-5$ \% & $5-10$ \% & $10-15$ \% & $15-20$ \% \\ \hline
$\Sigma_{p_{T}}$(\%) & & & & \\
(no SRC removal)
&$0.94^{+0.09}_{-0.10} {}^{+0.28}_{-0.26}$
&$1.04^{+0.09}_{-0.10} {}^{+0.17}_{-0.18}$
&$1.27^{+0.10}_{-0.11} {}^{+0.17}_{-0.14}$
&$1.53^{+0.16}_{-0.18} {}^{+0.19}_{-0.19}$\\
(with SRC removal)
&$0.70^{+0.12}_{-0.14} {}^{+0.28}_{-0.27}$
&$0.78^{+0.12}_{-0.14} {}^{+0.18}_{-0.18}$
&$0.99^{+0.12}_{-0.14} {}^{+0.18}_{-0.14}$
&$1.28^{+0.19}_{-0.22} {}^{+0.19}_{-0.19}$\\ \hline
$\Phi_{p_{T}}$ (MeV/$c$) & & & & \\
(no SRC removal)
& $4.0 \pm 0.8^{+1.4}_{-1.3}$
& $3.9 \pm 0.7^{+0.9}_{-1.0}$
& $5.0 \pm 0.8^{+0.9}_{-0.4}$
& $6.5 \pm 1.4^{+1.1}_{-1.4}$\\
(with SRC removal)
&$2.3 \pm 0.8^{+1.4}_{-1.4}$
&$2.2 \pm 0.7^{+0.9}_{-1.0}$
&$3.1 \pm 0.8^{+1.0}_{-0.4}$
&$4.6 \pm 1.4^{+1.1}_{-1.4}$\\ \hline
$n$ & 135075 & 161036 & 137261 & 45647\\
$\langle N \rangle$
& $129.22 \pm 0.05$& $110.20 \pm 0.04$&$92.22\pm0.04$&$82.19\pm0.07$\\ 
$\overline{p_{T}}$ (MeV/$c$)
& $434.95 \pm 0.07$ & $433.12 \pm 0.07$&$430.55\pm0.08$&$428.90\pm0.15$\\ 
$\sqrt{\overline{\Delta p_{T}^{2}}}$ (MeV/$c$)
& $284.26 \pm 0.07$ & $283.57 \pm 0.07$&$282.15\pm0.08$&$281.27\pm0.15$\\ 
$\sqrt{\langle \Delta M_{p_{T}}^{2} \rangle}$ (MeV/$c$)
& $25.340$ $\pm 0.006$ & $27.384$ $\pm 0.007$&$29.888$$\pm0.008$&$31.710$$\pm0.016$\\\hline
\tabularnewline[5pt]
\end{tabular}
\label{tab:ptsum80}
\end{table}

 \begin{table}
 \centering
\vglue0.2cm
\caption{Summary of mean $p_{T}$ fluctuations at 158~A~GeV/$c$.}
\vspace{0.25cm}
\tabcolsep=1pt
\begin{tabular}{ccccc} \hline
Centrality & $0-5$ \% & $5-10$ \% & $10-15$ \% & $15-20$ \% \\ \hline
$\Sigma_{p_{T}}$(\%) & & & & \\
(no SRC removal)
&$0.98 {}^{+0.07}_{-0.08} {}^{+0.28}_{-0.26}$
&$1.11 {}^{+0.07}_{-0.08} {}^{+0.14}_{-0.16}$
&$1.38 \pm 0.08^{+0.14}_{-0.14}$
&$1.45 \pm 0.10^{+0.17}_{-0.17}$\\
(with SRC removal)
&$0.78 {}^{+0.08}_{-0.09} {}^{+0.28}_{-0.26}$
&$0.90 {}^{+0.09}_{-0.10} {}^{+0.14}_{-0.17}$
&$1.12 {}^{+0.09}_{-0.10} {}^{+0.14}_{-0.14}$
&$1.16 {}^{+0.12}_{-0.13} {}^{+0.17}_{-0.17}$\\ \hline
$\Phi_{p_{T}}$ (MeV/$c$) & & & & \\
(no SRC removal) & $5.1 \pm 0.7^{+1.8}_{-1.6}$& $5.4 \pm 0.7^{+0.8}_{-1.3}$&$6.8\pm0.8^{+0.9}
_{-1.2}$&$6.5\pm0.9^{+0.8}_{-1.2}$\\
(with SRC removal) & $3.3 \pm 0.7^{+1.8}_{-1.6}$& $3.6 \pm 0.7^{+0.8}_{-1.4}$&$4.4\pm0.8^{+0.
9}_{-1.3}$&$4.1\pm0.9^{+0.8}_{-1.2}$\\ \hline
$n$ & 151191 & 153713 & 137414 & 102787\\
$\langle N \rangle$
 & $161.85 \pm 0.05$& $135.35 \pm 0.05$&$113.06\pm0.05$&$97.47\pm0.05$\\ 
$\overline{p_{T}}$ (MeV/$c$)
& $425.40 \pm 0.06$ & $423.76 \pm 0.06$&$421.70\pm0.07$&$419.49\pm0.09$\\ 
$\sqrt{\overline{\Delta p_{T}^{2}}}$ (MeV/$c$)
& $279.06 \pm 0.06$ & $278.30 \pm 0.06$&$267.23\pm0.07$&$276.18\pm0.09$\\ 
$\sqrt{\langle \Delta M_{p_{T}}^{2} \rangle}$ (MeV/$c$)
& $22.335$ $\pm 0.005$ & $24.379$ $\pm 0.005$&$26.723$$\pm0.007$&$28.612$$\pm0.009$\\\hline
\tabularnewline[5pt]
\end{tabular}
\label{tab:ptsum160}
\end{table}

\end{document}